\documentclass[preprint,pra,showpacs]{revtex4}
\usepackage{graphicx}
\begin{document}
\draft
\title{A possible resonance mechanism of earthquakes}
\author{V.V.Flambaum $^{1,2}$, B.S. Pavlov $^{2}$}
\address{$^1$School of Physics, University of New South Wales,
Sydney 2052, Australia}
\address{$^2$ New Zealand Institute for Advanced Study, Massey University, Private Bag 102 904, North Shore MSC 0745, Auckland, New Zealand.}
\date{\today}

\begin{abstract}
It had been observed by Linkov, Petrova and Osipov (1992) that there exist periodic  4-6  hours  pulses of $\sim 200 \,\mu$Hz
 seismogravitational oscillations ( SGO ) before 95 $\%$ of powerful earthquakes. We  explain  this  by  beating
between an oscillation eigenmode of a  whole tectonic  plate and  a  local eigenmode of an active zone. The beating  transfers
the  oscillation energy  from the remote zone of the  tectonic plate to the active zone, triggering the earthquake. Oscillation frequencies  of the plate  and  ones of the active zone  are tuned to a resonance by an additional  compression applied  to  the active zone due to  collision  of neighboring plates or  the  magma flow in the liquid underlay of  the  astenosphere ( the upper mantle). In the case when there are  three or more
SGO  with incommensurable  difference frequencies $\omega_m - \omega_n$  the  SGO beating pattern looks quasi-random, thus masking  the  non-random nature  of the  beating process. Nevertheless, we are  able to discuss a possibility of the  short term earthquakes predictions based on  an  accurate  monitoring   of  the beating  dynamics.
\keywords{Earthquake \and Resonance \and Beating}
\end{abstract}
\maketitle

\section{Qualitative picture}\label{sec:qual}
 Potential energy of an  elastic  object  is  usually defined  by an appropriate   quadratic form.
Enhancing of the quadratic form  implies  increasing  of the eigenfrequencies,  recall the  stretching  of a guitar string
 or bending  of a  relatively  thin  plate caused by  a normal force. Monitoring  of  frequencies of  SGO in connection
 with  elastic energy  storage  on the tectonic plate (in particular, due  to  bending under a normal force ) was  discussed in (Petrova and Pavlov 2008). On the  contrary, a tangential  compression  of an  elastic plate implies  decreasing  of  the eigenfrequencies, see  (Heisin 1967). While the lower eigenfrequencies of a  small tectonic  plate lie much  higher than  the lower  eigenfrequencies  of a similar  large plate, the tangential  compression   may bring the lower  eigenfrequency of the  small  plate in resonance with some  eigenfrequencies  of a large plate. Therefore, the oscillation energy of the corresponding mode  on the   large tectonic plate  may be  transferred  to the small plate in contact  ( or just  to an appropriate compressed  area - an active zone ) thanks to the beating phenomenon, which happens  between  two coupled oscillators with close frequencies. For instance, the dynamics of two coupled oscillators with almost equal eigenmodes frequencies  $\bar{\omega} \pm \omega_{\delta}$,  with coordinates $x$ and $y$ and interaction $ \delta xy$,  is described  in (Landau and Lifshitz1969),  as
   \[
  x =  A \cos (\omega_{\delta}t + \varphi  ) \cos(\bar{\omega}t  ),\,\,
  y =  B \sin (\omega_{\delta}t + \varphi  )  \cos(\bar{\omega}t  ).
  \]
Here the beating frequency $\omega_{\delta} << \bar{\omega}$.
This example illustrates periodic (with the frequency $\omega_{\delta}$) migration of energy from one plate to another.
General algebra of beating is reviewed in Appendix B. Further calculations  concerning the resonance
  interaction  of SGO  modes   for  contacting  small and large plates  requires deeper  mathematical analysis  and will be published by  V. Flambaum, G.Martin
   and  B. Pavlov  in a separate paper ``On   the resonance interaction of  seismogravitational  modes on tectonic plates" (in preparation). Here we present simple
   estimates.
 The frequency  of a transverse plane wave on a large thin plate depends on  the pressure  as  in  (Heisin 1967).
\begin{equation}
\omega =  2\pi \,\,\nu = k\,\,\, \sqrt{\frac{D k^2 - Q}{\rho h}},
\label{heisin}
\end{equation}
where $ Q$ is the  compressing tangential force  per unit length applied  along the direction of the motion of the wave, $\rho$ is the  density and  $h$ is the thickness  of the plate, $D = \frac{h^3 E}{12 (1-\sigma^2)}$, $ E$ is the Young's modulus, and $\sigma$ is the Poisson ratio. Solution for a spherical wave  is presented in appendix C.
We can use Eq. \ref{heisin} to  make a simple estimate of the effect of the compressing  tension.
 For the fundamental mode  on a  finite  plate the length  of the  wave vector is estimated  as $k \sim 1/L$, where  $L$ is the size of the plate.  Hereafter, for a typical  tectonic plate we  assume $E = 17.28 \times 10^{10}$ kg m$^{-1}$s$^{-2}$,
$\rho  = 3380 $ kg\,m$^{-3}$, $\sigma = 0.28$, and a  typical frequency of  a  large  tectonic plate vibration
 $\nu \sim $ 170 - 190   $\mu$ Hz, see (Petrova and Pavlov 2008).  This  allows  to estimate the   tangential  tension
 required  to bring  lower  eigenfrequencies  of  a small plate into resonance with lower  eigenfrequencies  of the  large plate.
Significant reduction of the frequency of  a small plate (which brings it to a resonance with a frequency of a large plate) happens near
the root  of  the expression under the square root  in Eq. (\ref{heisin}),  $ D k^2 = Q$. This gives us an estimate for the required pressure:
\begin{equation}
P \sim 10^{10} \left(\frac{h^2}{L^2}\right) \mbox{Pa}
\end{equation}
Hereafter we  assume that  the active zone is  a small plate  $\Omega$ which is in contact  with  a large plate $\Omega_t$.
 The typical  linear size of the small plate   is $L\sim $ 100 - 200  km.
Large tectonic plates extend  to  $L_t \sim $ 1000-10000 km and are  $ h  \sim $ 200 - 300 km thick  on the continents, but
much thinner, $h =$  30 - 100  km on  the oceans bottoms. A small ratio $h/L$ may reduce the required compression  for the resonance up to  two orders of magnitude. An upper estimate of an  existing  compression  may be given by a crushing pressure $P \sim  10^{9}$ Pa of  the  material composing  the  plates.

The resonance may also appear for a much smaller
 compression, if the tectonic plate is thick and the small  one (active zone) is relatively thin (e.g. under ocean).
  Indeed, for small $Q$ the  plate frequency is
 \[
 \omega \sim  (1 / L^2)\sqrt{(D/\rho h)} \propto h/L^2,\,\,  L^2/h \sim  L_t^2/(h_t n^2)
 \]
  There  may be also resonances between  the fundamental frequency  of the small plate and  the higher SGO modes of  the large tectonic  plate ($k L_t \sim n, \, n=1,2,... $ ). For small $Q$ the  tectonic plate frequency and the resonance condition are reduced  to
   \[
   \omega \sim (n^2  / L_t^2)\sqrt{(D/\rho h_t)},\,\,
   L^2/h \sim  L_t^2/(h_t n^2).
     \]
   Finally, there may be a resonance between the different types of the modes on different plates which have different fundamental frequencies (e.g. transverse, longitudinal and surface (Rayleigh) modes). The lowest mode on the small plate may resonate with a higher  mode on the large plate. The increasing compression  on the small plate in this case defines scanning of the frequency until it comes into a resonance with one of the frequencies on the large plate.
 An  efficient mechanism of the scanning may arise  also due to a bending of the  plates by the non-tangential  force. The bending increases  the  potential  energy of  the elastic deformation  and  the oscillation frequencies.
The change of frequencies of the tectonic plates due to the bending has been discussed, e.g.  in (Petrova and Pavlov 2008).
Moreover, this effect is seen in the data of observations presented below, see Fig. \ref{F:figure_0}.

Assume that  initially the  SGO  are registered  on a large plate.  If the small and  large plates  are disconnected, their  oscillations  are independent. The   compression   may bring  the  frequency of the  small plate in resonance with  some frequency of SGO  of the large plate. Then even  a   weak  interaction between  the SGO processes  on the plates  results in forming  the perturbed  SGO  mode of the  pair of  plates  $\Omega \cup\Omega_t$,  manifesting the beating  of the modes  and  causing  migration of  energy from one plate to another.  The sum of energies of
oscillations of the plates  remains constant.  In absence  of  exact resonance the  energy transfer is not complete, but
when the resonance  becomes sharper, the energy  transfer  becomes  fuller.  When the energy  of the  large plate  comes to the small plate, the  mean  amplitude $A$  of SGO  on the  small plate  becomes  enormous  due  to the energy  conservation law $A/A_t \sim L_t\sqrt{h_t}/ L \sqrt{h}, $ triggering  the earthquake.

Even a partial transition  of energy  of a resonance  SGO -mode from the  large tectonic plate  to the active zone  may cause an enormous effect. For instance  the elastic energy stored in a single  SGO mode with frequency 200 $\mu$HHz and amplitude   $2\times 10^{-3}$ m  on the tectonic plate with area  $ 10^{14}$ m$^2$, thickness 10$^5$ m and density 3380 kg m$^{-3}$ is estimated as $54\times 10^{9}$ joules, which is almost equivalent to the  4M  earthquake in Johannesburg (South Africa) November  18, 2013. Even a  small part  of  this amount may trigger a  powerful earthquake, and  should be taken into account  when  considering a realistic mechanism  of the earthquake. Estimation of the  transfer coefficient defining the   transfer  of energy  in course of resonance  beating  of  SGO modes  would probably  help to  develop more  realistic Earthquake theoretical scenario.

The above  resonance interpretation of the earthquake mechanism  may be useful for  short-term  earthquake predictions,
see section 3, Conclusion.  For instance,  if  there are  only two  interacting modes with  frequencies  $\omega,\omega_t$, the beating is periodic. If we have registered  two pulses manifesting the  moments when the migrating energy  is  accumulated  on the large plate, then  in a half period after  the second pulse  the whole  energy will  be  already on the small  plate, and may trigger the earthquake.  But even if  we  observed a single pulse  and registered the  moments of maximal and   the  previous  moment  of minimal energy (amplitude) in  the remote zone  of the large plate, we  are able to predict the  moment when  the energy is maximal  on the small plate. This moment is  coincident with the next   minimum  of  the  energy  on the  large plate.  Note that the observations may be done  in the remote zone of the large plate,  very far from the active  zone. The case  of  perturbation of a multiple eigenfrequency may be considered  based on an appropriate aperiodic  beating algebra, see  Conclusion, section 3.

{\bf Suggestion of experiment}. Mathematics can't yet  provide reliable results for resonating  eigenmodes  of  plates under various  (tangential  or/and  normal ) tensions, beating frequencies and transferred energy. However, a more reliable way to investigate these problems may be a laboratory experiment with  plates of different sizes and shapes.  For
  instance,  most  natural question is  one on existence and  structures  of    oscillation modes, localized
essentially on   a small  active zone $\Omega$  or  on the complement $\Omega_t$, with close  frequencies
 $\omega \sim  \omega_t$, yet   without  the   resonance condition imposed, $\omega \neq \omega_t$. For frequencies approaching the resonance  $\omega -\omega_t \to 0 $, the experiment  may  help  to  recover their dependence  on  tensions  applied   on   the active zone, and  give  essential data  for constructing  and  fitting  of the  mathematical model  of  the general  ( aperiodic ) scenario  of  beating  of  the perturbed multiple modes.

{\bf The whip effect}. The  beating phenomenon is  observed while  the   beating frequency   is  sufficiently  large  compared  with the speed of  change  the  frequencies, caused  by the  bending  of  the  large plate .  If  the  changes go  faster, then  possibly  the  resonance condition  is  satisfied  only  once  during  the  observation  period. Nevertheless  the  earthquake  may be triggered   in this case  too, by  an  analog  of  the celebrated   whip  effect, which manifests, due  to  the energy conservation, the  amplitude growth  while the  wave, running along the   thinning  channel, is approaching the  thin end (of the whip). Indeed,  that may happen on the ocean bottom, if the  tectonic plate is  gradually  thinning along the wave track.
Generally  a combination of  the  beating and  whip scenarios  of the  earthquake  is possible, while  few beating  are
terminated  by  the whip effect  triggering  the  earthquake  at  the moment of reaching  resonance.

 In this text we  neglect an important phenomenon of dissipation in  the SGO process.
 A preliminary  discussion of  this matter  may be found in (Ivlev et al. 2012).

\section{Observations and their resonance interpretation }\label{sec:obs}

 In  the paper  by  Linkov et al (1992) the  pulses  of SGO   were discussed as  typical precursors of   powerful earthquakes, arising with probability  95 \%. The  spectral nature of the SGO was  demonstrated in (Petrova and Pavlov 2008). Additional unpublished information was  kindly provided to us by L. Petrova, who provided us so-called  spectral-time cards  constructed by  herself based on monitoring of the  SGO process  preceding the  earthquake  26 September 2005  in Peru (see Fig.1). L. Petrova also attracted our attention to  some details on  the cards, which  may be  considered as  precursors of the earthquake, but   were not interpreted yet  properly. First, there are  two  ``pulses'' registered  on SSB station (France) in the zones  $\Delta^{SSB}_2 = (190,200)\times(55,65); \Delta^{SSB}_3 = (200,210)\times(145,165)$ introduced in the Appendix A
   below, separated by the time interval  96 hours.  We  believe  now, based on above  resonance interpretation,  that they are SGO -beating on the large plate  situated  on the  way  of the waves  coming from  Peru  location  to  the  SSB station in France.   Secondly,  she noticed  a  ``shock"  between the  pulses  at  $T = 87$  h, causing   generation of three oscillation modes, clearly  registered   on the  INU-station (Japan).
  We  guess, again based  on our resonance interpretation, that  the  corresponding  signal  may come from the small plate. Finally, the  earthquake succeeded at the moment T = 172 h, in 48 hours after the  second pulse, and  96  hours   after the  shock  at the moment $87$ h, see  details in Appendix A.
 Notice that  our guess  does not permit to  describe in detail the way of migration of energy  from  the  active zone in Peru
 to the  remote  zone  on the large plate $\Omega_t$, where the SSB station is  located. But this is typical  for  most  of  experiments
 with  resonance systems, where just ``some" interaction of  oscillators involved  is important.

The spectral-time cards also  give us  an evidence  of the bending effect which may tune frequencies of the  plates to the resonance. Firstly, we can extract  it from  presence, in the remote zone,  of  modes  with growing  (for  growing  bending) and  decreasing    (for relaxed   bending )  frequencies.  Indeed, they  are  easily  seen on  the spectral - time  cards  as   ridges  extending   from   the   left down  to right up  or - vice versa - from left up   to right down. We  may guess that
these  ridges arose  from  the  unperturbed modes  on the  big plates $\Omega_t$, which are  excited, on the perturbed
 background, by the shock at the moment  $T = 87$, and  affected  by  the  bending  or  relaxation  ( due to partial destruction ) of the large plate  under the {\it normal tension}. Secondly , we  guess  that  there exist also an  unperturbed  mode on the large plate,  with an almost  resonance frequency, which has  a ``knot" at  the  location  of  the active zone, so that  it's frequency $\omega_t$ is not affected  by the local bending at the location. We guess that  it stays  in resonance  with  the corresponding  mode  on the active  zone  $\omega_t - \omega \approx 0 $ , tuned  by  the  {\it tangential  tension}. We guess  that  it  forms the beating pattern, registered  on the  domains  $\Delta^{SSB}_2 = (190,200)\times(55,65); \Delta^{SSB}_3 = (200,210)\times(145,165)$  of the  SSB card.  A  minor shift  ``up''  of the resonance frequency $(190,200) \to (200,210) $  may be caused  by  the details of the  shape of the contact  between the  plates, which do not  provide an exact separation  of compressing and  normal forces on the contact.

\begin{figure}
[ht]
\begin{center}
 \includegraphics [width=3.3in] {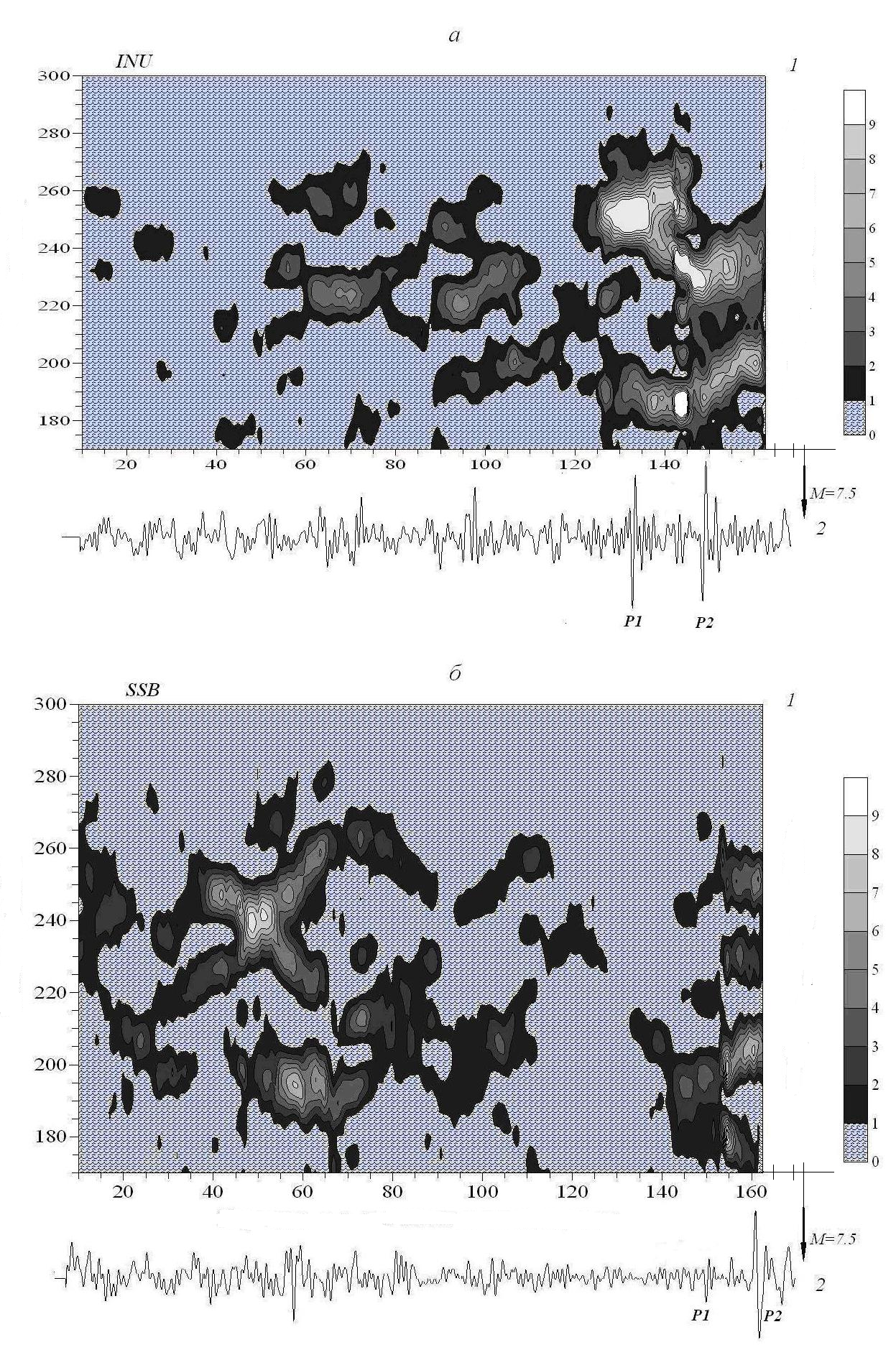}
\end{center}
\caption{Spectral-Time cards  constructed  by  L. Petrova  based  on seismo-gravitational oscillations recorded on INU  and SSB stations, in Japan  and in France respectively, during the period  18-26  September  2005 preceeding  the strong Eartquake  in North  Peru. The horizontal  axis  for time, is  graded in hours, the vertical axis , for the frequencies, is graded in  $\mu$Hz. }
\label{F:figure_0}
\end{figure}

\section{Conclusion} The  M8  Global test algorithm of the earthquake prediction, has been designed  in 1984 at the International Institute of Earhtquake Prediction  and Mathematical Geophysics (Moscow)  based  on the observation  that  almost 80 \% of  actual events  at the selected  location arise due to the stress built  up  thanks to  previous events at the   corresponding Earthquake-prone  (active)  zone.

 Though the algorithm was extremely efficient  providing  higher than 99 \% confidence level for  defined  Time intervals of Increased Probability (TIP) of the Earthquakes, yet some of  highly  dangerous events, like Tohoku  earthquake in Japan on  March 11  2011 were  not  predicted, because the black box  constructed based on the  M8 Global Test algorithm , removed the warning 70 days before the earthquake, see the retrospective analysis  of the Global Test effectiveness  in (Kossobokov 2013).

  We  did not plan, in our note, ``to come out with exact short term prediction algorithm''
but, inspired by  the author of  the above review, wish to provide our vision of the problem from the viewpoint of  mechanics and spectral theory.   We  hope that  our observations may attract attention of experimentalists to the  resonance phenomena  (beating) in the  SGO process  on  elastic  thin plates.

We  keep  in mind   that , in the  linear approximation of the elasticity  theory, the  seismogravitational oscillations  with the  frequency  circa  200 $\mu$Hz  and  the corresponding  pulsations as   precursors of the earthquakes,  correspond to  very long transverse waves. This  long waves  are not affected  by  local  shape  variations of  the plates (coarse  grain or  fine grain) and the relief. This  allows  to  model the  SGO  in the laboratory, clarifying  the basic questions of the SGO process, including the beating , the
    energy migration and  triggering  an earthquake.  This may help further  refining of the  M8 Global  Test algorithm, with regard of possible resonance effects and the energy migration, to  improve  the  TIP prediction results.

\begin{acknowledgements}
We are grateful to L. Petrova for providing Spectral-Time cards and valuable discussions.
We also thank B. Belinskiy, N. Morozov  and J.Wolfe for useful discussions.
This work is supported by the Australian Research Council.
\end{acknowledgements}

 \section { Appendix A: More about Spectral-Time cards}\label{sec:apA}

 The spectral - time cards are obtained from the seismograms  via averaging   of  the square amplitudes of the oscillations with  certain frequency  on the  systems of 20 hours  time - windows, selected by  shifting  an initial window  by  30 minutes on each  step. The seismograms underwent previously a double  filtration with  Potter  filter: the high frequency one, with boundary period 6 h, and one more, with  the window 1-6 h.   The boundaries of the   domains  on the cards,
where the averaged  amplitude  of the  SGO mode  with  certain frequency $\nu$, at the  given moment $T$  of time,
exceeds given  value $A$, form a system of isolines in the frequency/time  coordinates $\nu,T$.
   The  relief  of the  window  averaged  squared amplitude on the cards  is graded  by the isolines , with the step  $\delta A^2 =  \frac{1}{10} \left[ A^2_{max}  - A^2_{\min}\right]$, and  is painted  accordingly  between the isolines, with  dull grey for the  background  value  $A^2_{\min}$  of the squared amplitude  and white for the
maximal value  $ A^2_{max}$.
   The  results of monitoring represented on the cards
 correspond  to  the  trains  of SGO  with  growing $ a) $,  constant $ b)$ and decreasing  $c)$ frequency
 $\nu$\\
a) INU 88 $\longrightarrow$ 107, 220 $< \nu <$ 235, \\
b) SSB  50 $\longrightarrow$ 65, $ \nu \approx$ 195,\\
c) SSB  70 $\longrightarrow$  87, 265 $> \nu>$ 247,\\
and   brief ( 6-20 hours) stationary  SGO  modes with high amplitudes in  the  (conventionally) rectangular  zones
on the cards in  frequency - time  coordinates  as  $ \Delta \nu, \Delta t$
$\Delta^{SSB}_1 = (235,245)\times(45,55)\\
\Delta^{SSB}_2 = (190,200)\times(55,65)\\
\Delta^{SSB}_3 = (200,210)\times(145,165)\\$
$\Delta^{INU}_1 = (240,260)\times(125,139)\\
\Delta^{INU}_2 = (225,224)\times(142,150)\\
\Delta^{INU}_3 = (180,190)\times(142,145)\\$
 One can  see  on the SSB card  two  groups  of stationary modes with  almost equal frequencies and visually similar relief
in  $\Delta^{SSB}_2$ and  $\Delta^{SSB}_3$, which  were  identified  by   L. Petrova   as  ``seismo-gravitational
pulsations'', see  comments in previous section. Dr. Petrova  also attracted our attention  to  a family  of  prolonged  (up to 50 hours)  SGO trains with growing frequency
(1 $\mu$Hz/hour) on the intervals  $(50,160)^{INU}$ $(0,110)^{SSB}$  and  almost total absence  of the modes with growing frequency  on the complementary intervals  $(0,50)^{INU},(110, 145)^{SSB}$.

Vice versa, the  modes with  decreasing frequency are absent  on the
time interval  $(87, 120)$ on both cards. Dr. Petrova suggested that probably some  important event (a shock?)  succeeded at the moment
 $87$, which  excited  three  SGO modes  on the interval  $(87, 120)^{SSB}$, two of them also clearly seen  on   $(87, 120)^{INU}$, see our interpretation in previous section.
 The  extent  of  the  clearly seen  part  of the  middle train , measured  on the middle line  of the corresponding ``ridge'' on the
 interval  $(87, 120)^{SSB}$  is about 24 hours, and the  extents  of the upper and lower modes  are  longer and  shorter than the middle
 one by the  intervals  proportional to  the  difference  of frequencies  of the modes.

 \section{ Appendix B:   Algebra of beatings}\label{sec:apB}

 The  problem on   beating  of the seismogravitational modes  has a  simple
algebraic nature:  it is modeled  by   a system of coupled oscillators with multiple
eigenfrequency $p_0$ which is perturbed  such that  the multiple  eigenvalue is  split   into
 a  starlet $p^s_{\delta} = p_0  +  \delta \alpha^s$  under a
minor perturbation, which  also  transforms the  initial eigenbasis $\left\{ e_0^s \right\} \longrightarrow
\left\{ e_{\delta}^s \right\}\equiv U_{\delta} \left\{ e_0^s \right\} $, with an unitary generator which, in
simplest case , is represented   by an exponent of an antihermitian  matrix $ U_{\delta} \equiv exp [\delta B]$.
Both  the starlet and the generator of  rotation the basis are usually  found  in terms of  normal modes, and
the perturbed evolution  is  represented as  a  linear  combination of the normal modes
\[
u(t) = \sum_s A_{\delta}^s cos[( p_0  +  \delta \alpha^s)t + \varphi_s ] U_{\delta} e^s_0
\]
The  energy
 $E_{\delta} (u) = \frac{1}{2}\left[ \parallel u_t \parallel^2 +  \sum_s |A_{\delta}^s|^2 |p^s_{\delta}|^2 \right] $
 of  the  perturbed  evolution $u_{tt} + A_{\delta} u = 0$ and  the  unperturbed energy   of  the  unperturbed  evolution  $u_{tt} + A_{0} u = 0$
  $E_{0} (u) = \frac{1}{2}\left[ \parallel u_t \parallel^2 +  |p_{0}|^2 \parallel u \parallel^2\right]$  are conserved.
The  unperturbed  values of energy  $E_0 (P^s_0 u )$ of  the  projections $P^s_0 u (t) \equiv  e^s_0 \rangle \langle e^s_0 ,u (t)\rangle $ of the perturbed evolution onto the  eigenvectors  $e^s_0$ of the  unperturbed generator $A_0$, being averaged over properly selected time  window,
expose  the beating phenomenon parametrized  by the characteristics of the splitting starlet and the  eigenbasis rotation:\\
$\frac{1}{\Delta}\int_{T - \Delta/2}^{T + \Delta/2}  E_0 (P_s u (t)) dt$
$\approx \frac{p^2_0}{2}\sum_{n,m}
\bar{A}_{\delta}^n\, A_{\delta}^m \cos[\delta (\alpha_n - \alpha_m) t + \varphi_n - \varphi_m] \overline{\langle e^0_{s},e^{\delta}_{n} \rangle}\langle e^0_{s},e^{\delta}_{m} \rangle$,\\
which  looks quasi-random while the  difference  frequencies $\delta (\alpha_n - \alpha_m)\equiv \omega_m - \omega_n$
are  incommensurable.
In the case  when there are only  splitting of the  multiplicity  two $ p_0 \to  p_0 \pm \delta $ the beating is  periodic, with  the  difference frequency  $ \delta\pi $, thus allowing to  preview  arising  maximal  energy
values  of the migrating  energy  on the active zone, as  noticed  in the  end of  section \ref{sec:qual}. In  the case of  incommensurable difference frequencies  a similar  previewing remains possible  too, despite  a  quasi-random
character of the  beatings  dynamics.
 The case  of the tectonic plates  is reduced to  the above  case  of   the  connected  oscillators  via  considering
 the  boundary values  of the  solutions of the corresponding  biharmonic wave equation  $\rho u_{tt} + L u = 0$.  The  corresponding spectral problem
 requires  considering  the    boundary  form for the generator $L$ : $\langle L u, v \rangle - \langle u, Lv \rangle$  which is reduced , see  (Pavlov 2001) to  a
  boundary integral $\int_{
} [ \langle D\bar{ u}, N v \rangle  - \langle N \bar{u}, D v \rangle] d\Gamma  $  and  vanishes if  appropriate boundary conditions   with an  hermitian matrix $B$ are imposed:  $[ N u - B Du ]\big|_{\Gamma} = 0$  onto  the  boundary values  $N u , Du$.

 Essential simplification of the
   original   spectral problem is  obtained while the  Dirichlet-to-Neumann   map, see
(Pavlov 2001), transforming the  boundary values of the  homogeneous problem
 $ L u = \lambda \rho u$ $N u = {\mathcal{DN}}(\lambda) D u$   one to another is  substituted by an appropriate finite-dimensional  rational approximation
 \[
 {\mathcal{DN}}(\lambda)\,\,\, \longrightarrow   P_E {\mathcal{DN}}(\lambda) P_E \equiv \,\,\,{\mathcal{DN}}_{E}(\lambda) ,
 \]
 which corresponds to  substitution of the original problem by
 a corresponding  fitted  solvable model
(Pavlov et al 2010).
\section{ Appendix C:   Wave equation and separation of the variables}\label{sec:apC}

The  viscosity  of the liquid underlay  is small for  relatively slow movements, which correspond  to  the  frequency  200  $\mu$Hz and typical amplitudes  of the  SGO. Based on analysis of  a thin plate floating on a liquid underlay, we  eliminate the hydrodynamicla  variables  obtaining,  see (Chung and Fox 2009), after an appropriate renormalisation,  the  biharmonic wave  equation for the tranverse waves on the thin plate in the following form, see (Heisin 1967, Landau and Lifshitz 1970):
\begin{eqnarray}
\nonumber
\,\,\,\,\,\,\,\rho h v_{tt} + \beta \, v_t + D \Delta^2 v  + Q \Delta v = 0 \,\,\,\, \stackrel{v = e^{i\omega t} u }\longrightarrow \\
\label{spectral}
\longrightarrow   D \Delta^2 u + i\omega \beta u + Q \Delta u = \omega^2\,\rho h u,
\end{eqnarray}
Hereafter  we neglect the  liquid  friction $\beta \, u_t$. It   may be  eliminated  via  an exponential factor  $u \to exp (-\beta\,\, t/2) u$   and  a re-normalization of the  frequency  $\omega^2 \to \omega^2 - \frac{\beta^2}{4} $
, while $\omega^2 - \frac{\beta^2}{4} > 0 $. The dependence of the frequency of the plane wave on the wave vector is given by
\begin{equation}
\label{Heisin_frequency_estim}
\omega = 2\pi \,\,\nu = \sqrt{\frac{D k^4}{\rho h} - \frac{Q k^2}{\rho h}}.
\end{equation}
For a spherical plate the  sepration of the variables is possible if the constant pressure force  Q is applied in a spherically symmetric way. The wave equation can be factorized  as
\begin{eqnarray}
\nonumber
\left( \sqrt{D} \Delta  +   \frac{Q}{2  \sqrt{D}}  + \sqrt{ \omega^2 \rho h + \frac{Q^2}{4 D}  } \right) \times\\
\times \left ( -\sqrt{D}  \Delta - \frac{Q}{2 \sqrt{D}} +
 \sqrt{ \omega^2 \rho h + \frac{Q^2}{4 D}  }  \right) u= 0,
\end{eqnarray}
and thus  reduced to  a pair  of separate equations. Their  solutions
in the subspace $E$ of  functions, independent  on  the angular variable, represented   on the spherical plate  $0\leq r \leq L$
  via Bessel functions
\begin{eqnarray}
\nonumber
J_0 \left( [\frac{\omega \sqrt{\rho h}}{\sqrt{D}}]^{1/2} \left[\sqrt{1 + \frac{Q^2}{4 \omega^2 \rho h D}} + \frac{Q}{2 \omega \sqrt{\rho h}\sqrt{D}} \right]^{1/2} r \right),\\
\nonumber
\, I_0 \left( [\frac{\omega \sqrt{\rho h}}{\sqrt{D}}]^{1/2} \left[\sqrt{1 +  \frac{Q^2}{4 \omega^2 \rho h D}} - \frac{Q}{2 \omega \sqrt{\rho h}\sqrt{D}} \right]^{1/2} r \right),
\end{eqnarray}
 where $J_0 $ is the standard Bessel function, and $I_0$   \,\,\, is a modified Bessel function of an imaginary argument $I_0 (z) = J_0 (iz)$. These solutions are regular at $r=0$. However, large plate may be modelled by a circular plate with a circular hole in the centre. In this case we should add two other solutions, the Hankel  functions $H^1_0$ and the modified
 Hankel function $K_0$. One also may use a model where a small plate is a sector inside a circular large plate. Solutions on the large plate in this case are Bessel function with the index $p$ determined by the angular size of the missing sector $\phi=\pi/p$, $p<1$.

The   eigenfunctions of the  above  biharmonic spectral problem on the small plate  are obtained as a linear combinations of the
Bessel functions $J $ and $I$, and  the  eigenvalues are calculated,
depending  on  the  contracting tension $Q$,   by   the
substituting of the linear combination into  relevant
boundary conditions on the border $r = L$ of the plate. Comparison
of the eigenvalues with  eigenfrequencies  of the large plate
defines the  condition  of  resonances.

We postpone all relevant mathematical details to the oncoming publication  by  V. Flambaum, G.Martin
   and  B. Pavlov  ``On   the resonance interaction of  seismogravitational  modes on tectonic plates" (in preparation)

{\bf REFERENCES}\\

Chung H., Fox C (2009){\it A direct relationship between bending waves and transition conditions on floating plates } (2009) Wave motion 46, 468-479.

Heisin D.E. (1967) {\it Dynamics of the Ice cover.} In  Russian. Hydrometeorological Publishing House.
Leninrad,  215 pp.

 Ivlev,L., Martin, G., Pavlov, B., Petrova,L. (2012) {\it  A zero-range model for localized  boundary stress  on a tectonic plate  with dissipative boundary conditions}, Journal of mathematics-for-industry, {\bf 4} (2012-9), 141-153.

Kossobokov V.G. (2013) {\it Earthquake  prediction; 20 years of global experiment} Natural Hazards {\bf 69} 1155-1177.

Landau L.D. , Lifshitz  E.M.  (1969) Mechanics, ( Volume 1 of A Course of Theoretical Physics ) Pergamon Press, Oxford, 221 pp.

  Landau L.D. , Lifshitz  E.M. (1970) Theory of Elasticity ( Volume 7 of A Course of Theoretical Physics ) Pergamon Press, Oxford, 218 pp.

  Linkov E.M. , Petrova L.N., Osipov K.S. (1992),
{\it Seismogravitational pulsation perturbatios in  the Earth  crust and in the atmosphete as  possible precursors of  powerful earthquakes} In Russian.
Doklady Academii Nauk, Physics of Earth,
{\bf 313}, 23-25.

 Martin G., Pavlov B., Yafyasov A.(2010).
{\it Resonance one-body scattering on a junction}
  Nanosystems: Physics, Chemistry, Mathematics,{\bf 1}, 108-147.

 Pavlov B.(2001). {\it S-Matrix and Dirichlet-to-Neumann  Operators}
In: Encyclopedia of Scattering, ed. R. Pike, P. Sabatier, Academic
Press, Harcourt Science and  Tech. Company, pp 1678-1688

 Petrova L., Pavlov B.(2008).
  {\it Tectonic plate under a localized boundary stress:fitting of a zero-range solvable model.}
 Journal of Physics A {\bf 41},   085206, 15 pp.\\

\end{document}